\documentclass[journal]{ijiss}
\usepackage[utf8]{inputenc}
\usepackage{listings}
\usepackage{enumerate}
\usepackage{amsmath,amssymb}
\usepackage{braket}

\ifCLASSINFOpdf
\usepackage[pdftex]{graphicx}
\else
\fi
\begin{document}
%
\title{Security of NEQR Quantum Image by Using Quantum Fourier Transform with Blind Trent}
%
%
%
%

\author{Engin~\c{S}AH\.{I}N*, \.{I}hsan~YILMAZ**
\address{*~Department of Computer and Instructional Technologies Education, Faculty of Education, Çanakkale Onsekiz Mart University, Çanakkale, Turkey.
e-mail: enginsahin@comu.edu.tr\\{**}~Department of Computer and Instructional Technologies Education, Faculty of Education, Çanakkale Onsekiz Mart University, Çanakkale, Turkey. e-mail: iyilmaz@comu.edu.tr}
}

%
%

\markboth{INTERNATIONAL JOURNAL OF INFORMATION SECURITY SCIENCE}{E. \c{S}AH\.{I}N and \.{I}.YILMAZ.}%
%



\ijisstitleabstractindextext{%
\begin{abstract}
In this study, the security of Novel Enhanced Quantum Representation (NEQR) of quantum images  are suggested by using the Quantum Fourier Transform ($QFT$) with blind trent. In the protocol, $QFT$ and keys are used to share signature with recipients.  So all members know only their signature information which are encrypted output of the $QFT$. This improves the security of the protocol. In addition, the security of the protocol is provided by using reorder $QFT$ output qubits with permutation of the blind trent. The security analysis expresses  security of the transfer of the image with effective secret key usage.

\end{abstract}

\begin{ijisskeywords}
Quantum Image Encryption and Authentication, Quantum Fourier Transform, Quantum Digital Signature.
\end{ijisskeywords}}

\maketitle



%

\section{Introduction}
%
%
\label{sec.1}
Conventional image security techniques, such as digital watermarking and image encryption, are commonly used to protect the image effectively. Possible threats in image transmission are capturing of image by unauthorized persons and authentication of the image. These issues are very important for special applications such as transmission of military satellite images or patient records.

There are many quantum image representations such as the "Qubit Lattice" representation for encoding a quantum image \cite{venegas1}, the "Real KET" representation \cite{latorre1}, Flexible Representation of Quantum Image (FRQI) \cite{Le2011},  Multi-Channel Representation for Quantum Images (MCQI) \cite{Sun2013}, Novel Enhanced Quantum Representation (NEQR) \cite{Zhang2013a}, Quantum Image Representation for Log-Polar Images (QUALPI) \cite{Zhang2013b}, Simple Quantum Representation of Infrared Images (SQR) \cite{Yuan2014}, Quantum States for M Colors and N Coordinates (QSMC \& QSNC) \cite{Li2013}, Normal Arbitrary Quantum Superposition State (NAQSS) \cite{Li2014} and Quantum RGB Multi-Channel Representation (QMCR) \cite{Abdolmaleky2017}. 

Also, in the literature there some studies on the security of quantum images. Iliyasu et. al. \cite{Iliyasu2010} proposed the WaQI scheme based on restricted geometric transformations for quantum image watermarking and authentication. Iliyasu et. al. \cite{ILIYASU212} proposed the improved WaQI scheme which is secure, keyless, blind and perfectly usable for authentication of the image owner. Later, Iliyasu et. al. \cite{Iliyasu2013} proposed GWaQI scheme with two-tier for greyscale quantum image watermarking and recovery. Yan et. al. \cite{Yan2015} proposed MC\_WAQI model based on MCQI representation for color quantum images.

In this study, we propose the protocol based on the $QFT$ with blind Trent for image security and authentication. The paper can be outlined as follows; in Section~\ref{sec.2},basic concepts of $QFT$ and $NEQR$ are explained; In Section~\ref{sec.3}, base stages of the protocol based on $QFT$ with blind trent are introduced. In Section~\ref{sec.4}, the security analysis of the protocol according to forgery and repudiation concepts are given. In Section~\ref{sec.5}, in the conclusion, some results are discussed.

%
%
%
%


\section{Related Works}
\label{sec.2}
\subsection{Quantum Fourier Transform}
\label{sec.2.1}
\begin{figure}[ht] 	
\centering
	\includegraphics{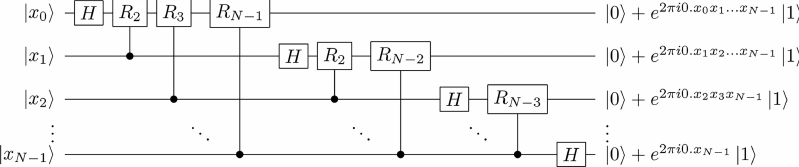}	
	\caption{Quantum Fourier Transform Circuit \cite{Nielsen2011}}
	\label{fig.1}
\end{figure}

Quantum Fourier transform is a application of classical discrete Fourier transform to the quantum states \cite{Nielsen2011}. 
The $QFT$ transform of an orthonormal basis set $\ket{0},\ket{1},...,\ket{N-1}$ can be defined as follows \cite{Nielsen2011}:
\begin{equation}\label{Eq.1}
\ket{x} = \frac{1}{\sqrt{N}}\sum_{y=0}^{N-1}e^{2\pi i x y/N}\ket{y}
\end{equation}
Where $N=2^n$ and orthonormal basis set is $\ket{0},\ket{1},...,\ket{2^n-1}$. The $\ket{x}$ state can be written in binary form as $x= x_0x_1...x_{N-1}$. The circuit of Quantum Fourier Transform for $\ket{x}$ can be seen in Fig.\ref{fig.1}. The $\ket{x}$ state is transformed into the phase of qubits as results of the $QFT$ transform.

\subsection{NEQR}
\label{sec.2.2}
NEQR uses the basis states of qubit sequence to store the color values. it represents grayscale images. Suppose the gray scale of images is $2^q$. Binary sequence $C^0_{YX}C^1_{YX}...C^{q-2}_{YX}C^{q-1}_{YX}$ encodes the grayscale value $f(Y,X)$ of the corresponding pixel $(Y,X)$ as follows \cite{Zhang2013a}:
\begin{equation}
\label{Eq.2}
\begin{split}
f\left(Y,X\right) = C^0_{YX}C^1_{YX}...C^{q-2}_{YX}C^{q-1}_{YX},\\ 
\quad C^{k}_{YX}\in \left[0,1\right], \quad f\left(Y,X\right)\in \left[0,2^q-1\right]
\end{split}
\end{equation}
\par
The representative expression of a quantum image for a $2^n \times 2^n$ image can be written as follows \cite{Zhang2013a}:
\begin{equation}
\label{Eq.3}
\begin{split}
\ket{I}=\frac{1}{2^n}\sum_{Y=0}^{2^{n}-1}\sum_{X=0}^{2^{n}-1}\ket{f\left(Y,X\right)}\ket{YX}\\=\frac{1}{2^n}\sum_{Y=0}^{2^{n}-1}\sum_{X=0}^{2^{n}-1} \bigotimes_{i=0}^{q-1}\ket{C^{i}_{YX}} \ket{YX}
\end{split}
\end{equation}

\section{The Protocol based on QFT with Blind Trent for NEQR Image}
\label{sec.3}
The participants of the protocol are Alice, Bob and Trent. Alice would like to send $2^q$ grayscale $2^n \times 2^n$ NEQR quantum image $\ket{I}$ to Bob by encrypting.In this case $N=q+2n$. Blind Trent is assumed as a manager of the protocol. Blind Trent manages some communication to provide security of the protocol. Bob can obtain and verify the image $\ket{I}$ with help of the blind Trent. 
\par
The protocol can be described with following phases.
\begin{figure}[ht]
	\includegraphics{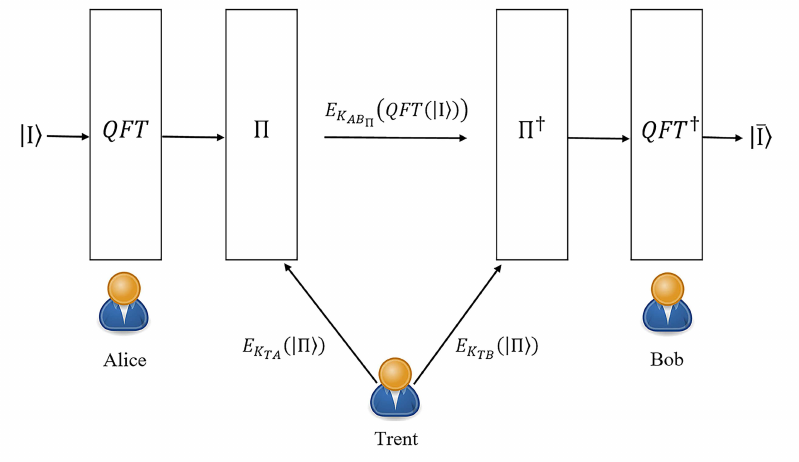}
	\caption{Security of NEQR quantum images based on QFT with blind Trent}
	\label{fig.2}
\end{figure}

\subsection{Initialization Phase}
\begin{enumerate}[(i)]	
\item Alice shares secret key $K_{AB}$ with Bob. Also blind Trent shares secret key $K_{TA}$ with Alice and secret key $K_{TB}$ with Bob. Participant's secret keys $K_{AB},K_{TA}, K_{TB}$  are shared by using quantum key distribution(QKD) protocol \cite{Bennett1984, Ekert1994}. The secret keys are used to encrypt quantum image to prevent any attackers. The encryption algorithm is given in Eq.~\ref{Eq.10} and Eq.~\ref{Eq.11}. The length of the all keys are $|K|=8N$. $K_{AB},K_{TA}, K_{TB}$ secret keys are created only once. Then  the secret keys can be divided into $8$-bit pieces. Each piece of the secret keys will be applied to every qubit of the QFT output according to the permutation determined by the blind trend. Different $K_{AB},K_{TA}, K_{TB}$ secret keys are created for each different sessions. If we apply stages in \cite{Yilmaz2017} to the NEQR image and use secret keys, we get the following stages.

\item Alice prepares her $2^q$ grayscale $2^n \times 2^n$ image $I$ with NEQR model . We assume that the length of the $\ket{I}$ is $|\ket{I}| = N$.	

\begin{equation}
\label{Eq.4}
\begin{split}
\ket{I} = & \frac{1}{2^n}\sum_{Y=0}^{2^{n}-1} \sum_{X=0}^{2^{n}-1} \ket{f\left(Y,X\right)}\ket{YX} \\
= &  \frac{1}{2^n}\sum_{Y=0}^{2^{n}-1} \sum_{X=0}^{2^{n}-1} \bigotimes_{i=0}^{q-1}\ket{C^{i}_{YX}} \bigotimes_{j=0}^{n-1}\ket{y^{j}_{Y}} \bigotimes_{j=0}^{n-1}\ket{x^{j}_{X}} 
\end{split}
\end{equation}
Where $\ket{C^{i}_{YX}}, \ket{y^{j}_{Y}}, \ket{x^{j}_{X}} \in \{\ket{0},  \ket{1}\}$. 

\item Blind Trent creates a permutation $P$ of a set of $\{1,2,...,N\}$ as follows \cite{Yilmaz2017}:

\begin{equation}
\label{Eq.5}
P = 
\begin{bmatrix}
1 & 2 & ... & N \\
P(1) & P(2) & ... & P(N)
\end{bmatrix}
\end{equation}

$P(i)$ can be expressed in binary string as follows:
\begin{equation}
\label{Eq.6}
\begin{split}
P_{binary}(i) =& P^0(i)P^1(i)...P^{k-1}(i) \\ 
&P^{j}(i) \in [0,1]
\end{split}
\end{equation}
Where $k=\log_2(P(i))$. Then Blind Trent prepares a quantum state $\ket{P(i)}$ from $P_{binary}(i)$ using by computational bases (0$\rightarrow$$\ket{0}$ and 1$\rightarrow$$\ket{1}$). Then blind Trent prepares a quantum state of $\ket{P}$ as follows:
\begin{equation}
\label{Eq.7}
\ket{P} = \bigotimes_{i=1}^{N}\ket{P(i)}
\end{equation}

Blind Trent creates encrypted versions of that permutation as follows:
\begin{eqnarray}
\label{Eq.8}
\ket{P_A}  = E_{K_{TA}}(\ket{P}) \\
\label{Eq.9}
\ket{P_B} = E_{K_{TB}}(\ket{P})
\end{eqnarray}
$E_K(.)$ is a quantum one-time pad encryption algorithm which is firstly defined by Kim et. al. \cite{Kim2015} and used by Yılmaz \cite{Yilmaz2017} to improve security of the protocol against forgery attacks. To further improve the security of this quantum encryption algorithm, we reorganized the algorithm of Zhang et. al. \cite{Zhang2016} as follows:

\begin{equation}
\label{Eq.10}
\begin{split}
\scriptscriptstyle
E_K(\ket{I})=\bigotimes_{i=0}^{N-1} {\sigma_{x}}^{K_{8i}} {\sigma_{z}}^{K_{8i+1}} T {\sigma_{x}}^{K_{8i+2}} {\sigma_{z}}^{K_{8i+3}}\\ 
\scriptscriptstyle
T {\sigma_{x}}^{K_{8i+4}} {\sigma_{z}}^{K_{8i+5}} T {\sigma_{x}}^{K_{8i+6}} {\sigma_{z}}^{K_{8i+7}}\ket{I_i}
\end{split}
\end{equation}
\begin{equation}
\label{Eq.11}
T=\frac{i}{\sqrt{3}}(\sigma_{x}-\sigma_{y}+\sigma_{z})
\end{equation}
Due to using $T$, encrypted message cannot be forged \cite{Kim2015}. Where the key length is $|K|=8N$. Then, blind Trent sends encrypted $\ket{P_A}$ to Alice via quantum channel.

\item Alice decrypts $\ket{P_A}$ and makes measurement to obtain $P$ \cite{Yilmaz2017}.

\item Alice applies $QFT$ to NEQR image and obtains following state:

\begin{equation}
\label{Eq.12}
\begin{split}
&QFT\left(\ket{I}\right)=\frac{1}{2^n}\sum_{Y=0}^{2^{n}-1} \sum_{X=0}^{2^{n}-1}\\
&\left(QFT\left(\ket{c^{0}_{YX}...c^{q-1}_{YX}y^{0}_{YX}...y^{n-1}_{YX}x^{0}_{YX}...x^{n-1}_{YX}}\right)\right) \\ 
&=\frac{1}{2^n}\frac{1}{\sqrt{2^{N}}}\sum_{Y=0}^{2^{n}-1}\sum_{X=0}^{2^{n}-1}\left(\ket{0}+e^{2\pi0.x^{n-1}_{YX}}\ket{1}\right)\\
&\otimes \left(\ket{0}+e^{2\pi 0.x^{n-2}_{YX}x^{n-1}_{YX}}\ket{1}\right)\otimes...\\
&\otimes\left(\ket{0}+e^{2\pi0.c^{0}_{YX}c^{1}_{YX}...x^{n-1}_{YX}}\ket{1}\right) 
\end{split}
\end{equation}
Where $N=q+2n$. Also we show it as follows:
\begin{equation}\nonumber
\begin{split}
\bigotimes_{i=0}^{N-1}\ket{I_i}=&\frac{1}{2^n}\sum_{Y=0}^{2^{n}-1} \sum_{X=0}^{2^{n}-1} \bigotimes_{i=0}^{q-1} \ket{C^{i}_{YX}} \bigotimes_{j=0}^{n-1} \ket{y^{j}_{Y}}\\ 
&\bigotimes_{j=0}^{n-1}\ket{x^{j}_{X}}
\end{split}
\end{equation}
\begin{equation}\nonumber
QFT\left(\ket{I}\right) = \frac{1}{\sqrt{2^N}}\bigotimes_{i=0}^{N-1}QFT\left(\ket{I_i}\right)
\end{equation}

\end{enumerate}

\subsection{Signing Phase}
\begin{enumerate}[(i)]
\item Alice reorder $QFT\left(\ket{I_i}\right)$ states with permutation of blind Trent.
\begin{equation}
\label{Eq.13}
\begin{split}
\ket{A(Q)} = &SWAP(P(i))(QFT(\ket{I_i}))\\
&i=0...N-1
\end{split}
\end{equation} 

\item Alice encrypts all qubits of  Eq.~\ref{Eq.13} with secret key $K_{AB}$.
\begin{equation}
\label{Eq.14}
\begin{split}
\ket{A(S)} = &E_{K_{AB_{P_A(i)}}}(\ket{A(Q)})\\
&i=0...N-1
\end{split}
\end{equation} 

\item Alice sends $\ket{A(S)}$ to Bob via quantum channel. 
	
\item Alice encrypts $QFT(\ket{I_i)}$ with secret key $K_{AB}$ with the encryption algorithm \cite{Yilmaz2017} (see references there in) and sends to blind Trent via quantum channel.
\begin{equation}
\label{Eq.15}
\ket{AT(S)} = E_{K_{AB}}(QFT(\ket{I_i}))
\end{equation} 

\item Blind Trent encrypts the $\ket{AT(S)}$ with the secret key $K_{TB}$.
\begin{equation}
\label{Eq.16}
\ket{TB(S)} = E_{K_{TB}}(\ket{AT(S)})
\end{equation} 

Then, blind Trent sends the above encrypted state to Bob via quantum channel.	
\end{enumerate}

\subsection{Verification Phase}
\begin{enumerate}[(i)]
\item Bob decrypts $\ket{TB(S)}$ with the secret key $K_{BT}$ and gets $\ket{AT(S)}$. Then, Bob decrypts the states $\ket{AT(S)}$ with $K_{AB}$ and gets $QFT(\ket{I_i})$ states. Bob applies $QFT^{-1}$ and measures the $\ket{I_i}$ states as the stage of image retrieving in \cite{Zhang2013a} and saves the results as $\tilde{I}$.

\item Bob decrypts $\ket{A(S)}$ by using secret key $K_{AB}$ and obtains $\ket{A(Q)}$.

\item Bob asks to blind Trent for permutation. Then, blind Trent sends $\ket{P_B}$ to Bob via quantum channel. Bob decrypts $\ket{P_B}$, makes measurement and obtains $P$. Bob reorder $\ket{A(Q)}$ states with permutation of blind Trent and obtains $QFT(\ket{I_i})$.
\begin{equation}
\label{Eq.17}
\begin{split}
QFT(\ket{I_i}) = &SWAP(P(i))(\ket{A(Q)})\\
&i=0...N-1
\end{split}
\end{equation}

\item Bob applies $QFT^{-1}$ and gets $\ket{I_i}$, then makes measurement onto that states as stage of the image retrieving in \cite{Zhang2013a} and obtains $\bar{I}$. Bob checks equality of $\tilde{I}$ and $\bar{I}$. If $\tilde{I}=\bar{I}$, Bob will announce that the signature is valid, otherwise the signature is rejected and the protocol is aborted. If the image is valid, then Bob encrypts the valid image $I$ with encryption algorithm.
\begin{equation}
\label{Eq.18}
\ket{BT(S)} = E_{K_{BT}}(\ket{I_i})
\end{equation} 

Then Bob sends $\ket{BT(S)}$ to blind Trent.

\item Blind Trent decrypts $\ket{BT(S)}$ with secret key $K_{BT}$ and measures the states as the stage of image retrieving in  \cite{Zhang2013a} and obtains $\bar{I}$.

\item Blind Trent also asks Alice for sending $I$ to him. Alice encrypts the valid image $I$ with encryption algorithm.
\begin{equation}
\label{Eq.19}
\ket{AT(S)} = E_{K_{AT}}(\ket{I_i})
\end{equation} 

Then Alice sends $\ket{AT(S)}$ to blind Trent.

\item Trent decrypts $\ket{AT(S)}$ with secret key $K_{AT}$ and measures the states as the stage of image retrieving in \cite{Zhang2013a} and obtains $\tilde{I}$. Blind Trent checks the equality of the $\tilde{I}$ and $\bar{I}$. If they are equal  then stores the image $I$ with Alice's identification for later traceability.

\end{enumerate}

\section{Security Analysis}
\label{sec.4}
Main requirements of the quantum digital signature protocols to provide unconditionally security are that the signature should not be repudiated by the signatory, and any attacker cannot forgery signatory's signature.

\subsection{Impossibility of Forgery}
Firstly, we consider insider attacker for the protocol. We assume that Bob is illegal participant and wants to create a signature of Alice. Even if Bob knows the details of the signature protocol he cannot create Alice's signature because of blind Trent.  Bob cannot create Alice's signature without knowledge of blind Trent. After the end of the legal signature protocol, Bob may change correct image $I$ to $\bar{I}$. Because of the knowledge about correct $I$ of blind Trent, Bob cannot achieve forgery. 

Secondly, any attacker may try to forge Alice's signature. Any attacker cannot achieve forgery because the states in the results of the $QFT(\ket{I_i})$ can be reordered with blind Trent's permutation to produce a correct signature of Alice. Even if any attacker can get the permutation, the permutation will be changed by blind Trent for every signature session. Blind Trent must be part of the protocol. So any attacker cannot achieve collective forgery. Further, any attacker may change $QFT(\ket{I_i})$ state by applying unitary transformation. Then, Bob and blind Trent can decide changed state by comparing $I$ and $\bar{I}$.

\subsection{Impossibility of Repudiation}
In the protocol, Alice and Bob cannot repudiate the signature because of the management of protocol by blind Trent. Blind Trent controls some communication steps of the protocol. If Alice can send different $\tilde{\ket{I}}$ to the blind Trent and claim that the signature is not mine. Blind Trent can check the equality of the $\tilde{\ket{I}}$ from Alice and $\bar{\ket{I}}$ from Bob. Blind Trent can decide whether the signature protocol is valid or not.

\section{Conclusion} 
\label{sec.5}
Security of NEQR Quantum Image by Using Quantum Fourier Transform with Blind Trent is suggested. Alice expresses the image $\ket{I}$ into phase-space by using $QFT$. So the image $\ket{I}$ is expressed in phases of the output qubits of $QFT$. This improves the image security. Alice changes order of the output qubits of the $QFT$ according to permutation information which is sent by blind Trent. So any attacker does not know order of the qubits and also they cannot create a valid signature. 

In the encrypting algorithm, the length of keys is increased to the 8-bit to improve the security.  Any information (classical or quantum) in the protocol is sent by using encryption algorithm which is robust against forgery by insider/outsider attacker. Furthermore, decoy states can be used to be aware of Eve.

Bob can verify validity of the signature by the help of the blind Trent. Blind Trent must send the order of the qubits to the Bob to obtain real message $\ket{I}$ by using $QFT^{-1}$.       

The above security analysis implies that given protocol based on $QFT$ provides unconditionally security. In addition, our protocol provides higher efficiency, effective secret key usage and security for the transfer of NEQR images.
\ifCLASSOPTIONcaptionsoff
  \newpage
\fi

\end{document}